\documentclass[english,prl,reprint]{revtex4-2}
\usepackage[T1]{fontenc}
\usepackage[latin9]{inputenc}
\usepackage{color}
\usepackage{babel}
\usepackage{mathtools}
\usepackage{amsmath}
\usepackage{amssymb}
\usepackage{graphicx}
\usepackage[unicode=true,pdfusetitle,
 bookmarks=true,bookmarksnumbered=false,bookmarksopen=false,
 breaklinks=true,pdfborder={0 0 1},backref=false,colorlinks=true]
 {hyperref}
\hypersetup{
 pdfborderstyle=,citecolor=blue}

\makeatletter

\usepackage{color}
\usepackage{babel}

\usepackage{color}

\usepackage{color}
\usepackage{times}

\usepackage[hang]{footmisc}
\makeatother

\begin{document}
\title{Statistical Uncertainty Principle in Stochastic Dynamics}
\author{Ying-Jen Yang}
\email{ying-jen.yang@stonybrook.edu}

\affiliation{Laufer Center for Physical and Quantitative Biology, State University
of New York, Stony Brook, New York 11794, USA}
\author{Hong Qian}
\affiliation{Department of Applied Mathematics, University of Washington, Seattle,
Washington 98195, USA}
\date{\today}
\begin{abstract}
Maximum entropy principle identifies forces conjugated to observables
and the thermodynamic relations between them, independent upon their
underlying mechanistic details. For data about state distributions
or transition statistics, the principle can be derived from limit
theorems of infinite data sampling. This derivation reveals its empirical
origin and clarify the meaning of applying it to large but finite
data. We derive an uncertainty principle for the statistical variations
of the observables and the inferred forces. We use a toy model for
molecular motor as an example.
\end{abstract}
\maketitle
Thermodynamics has been the guiding empirical principle for statistical
physicists to understand heat engines and condensed matters \citep{landau_statistical_1980,huang_statistical_1987}.
Importantly, it identifies entropic forces conjugated to the observables
of interest and dictates force-observable relations such as the equation
of state and the Maxwell relations. However, textbook thermodynamics
was limited to describing state distributions in large, mechanical
systems at equilibrium. Extensions were thus needed for its application
to \emph{nonequilibrium }\citep{nicolis_self-organization_1977},\emph{
small} \citep{hill_thermodynamics_2013,bedeaux_nanothermodynamics_2020}, and
\emph{dynamical} \citep{jaynes_minimum_1980,presse_principles_2013}
systems.

The complexity of biological systems poses an additional challenge
for formulating such a theory of forces. The ``constituting individuals''
of complex biological systems, \emph{e.g.} a cell in a tissue or an
organism in a ecosystem, are themselves high dimensional. When describing
these complicated biological systems, it is often not practical to model
its constituting individuals from classical or quantum mechanics.
We consider these systems \emph{far from mechanics }as the practical
mathematical models that describe them are not Hamiltonian-based.
To formulate a thermodynamic theory of forces for biological systems,
we must formulate a theory that applies to not just nonequilibrium,
small, dynamical systems, but also systems that are \emph{not} modeled
by \emph{mechanics} \citep{szilard_uber_1929,mandelbrot_derivation_1964}.

The formulation of such a theory, known as the Maximum Entropy principle
(MaxEnt), has been summarized nicely by E. T. Jaynes \citep{jaynes_information_1957,jaynes_minimum_1980}
and put into practices \citep{presse_principles_2013}. This paper
mainly adds on two points. First, we revisit and advocate the empirical
origin of MaxEnt based on limit theorems in the idealization of having
infinite data. Second, closely following the logic of the first point,
we derive and explain the uncertainty principle between the statistical
variations of \emph{dynamical} observables and the conjugated path
entropic forces they infer.

We first argue that both of the two other mainstream derivations of
MaxEnt implicitly assume the idealized limit of infinite data when
MaxEnt is applied to real data. Specifically, both Jaynes' ``maximum-ignorance''
argument \citep{jaynes_information_1957} and Shore and Johnson's
axiomatic derivations \citep{shore_axiomatic_1980} are formulated
about the expected values of the observables. And, as true expected values
are only available in the \emph{data infinitus limit}, these formulations implicitly assumed data infinitum. Whenever
one measures a sample average from large but finite data and consider
the sample average a good approximation to the true expected value,
it is plugged into MaxEnt as if the data were infinitely big.

With this, we advocate the empirical derivation of MaxEnt based solely
on mathematical limit theorems of the \emph{data infinitus limit.
}This derivation has at least two advantages. On the one hand, it
explicitly states the data infinitum assumption and clarifies how
MaxEnt is used in finite but large data: MaxEnt applies to Big Data
as a leading order approximation, as how textbook thermodynamics is
applied to finite but large system. On the other hand, this derivation
shows three equivalent interpretations of the MaxEnt posterior with
clear connections to Bayesian conditioning. Further, it provides the entropy
function a statistical meaning instead of treating it as an auxiliary
function for inference.

We then revisit the limit theorem based derivation of the dynamic extension of MaxEnt \citep{csiszar_conditional_1987,chetrite_nonequilibrium_2013,chetrite_nonequilibrium_2015}, now commonly known as the
maximum caliber principle (MaxCal) \citep{jaynes_minimum_1980,presse_principles_2013}.
We use it to derive the uncertainty principle of the statistical variations of observables and forces, which we shall call it the Statistical Uncertainty Principle (SUP) for stochastic dynamics.
We will use a simple three-state toy model of molecular motor and
the data it produces as an example. The SUP is different from the
recently-celebrated thermodynamic uncertainty relation in stochastic
thermodynamics \citep{barato_thermodynamic_2015,horowitz_thermodynamic_2020}.
Our SUP is closer to the uncertainty principle in quantum mechanics
as both of them are from invertible mathematical transforms: Legendre
for SUP and Fourier for quantum.

\paragraph{Maximum Entropy Principle for Markov Processes \label{sec:Maximum-Entropy-Principle-for-processes}}

The empirical derivation of MaxEnt for state distribution data of independent and identically-distributed
(i.i.d.) ensemble  has been revisited by one of
us \citep{qian_counting_2020,lu_emergence_2022}. Here, we briefly
revisit the data-driven derivation of its extension to correlated
data about transitions \citep{csiszar_conditional_1987,chetrite_nonequilibrium_2013,chetrite_nonequilibrium_2015}.

Before we begin, let us first remark that applying MaxEnt to stochastic
processes is conceptually straightforward from either Jaynes' argument
of least-bias \citep{shore_axiomatic_1980} or Shore and Johnson's
axioms \citep{shore_axiomatic_1980,presse_principles_2013}: one simply
replaces state distribution with path distribution. Jaynes called
this the Maximum Caliber principle (MaxCal) \citep{jaynes_minimum_1980}.
In this generalization, the stochastic process needs not to be Markovian or have
a steady state. Once we know the expected values of some path observables,
it can be used in MaxCal to provide an update on the our model of
path probabilities. However, if one aims to get these expected values
from data, one has to rely on the law of large number for convergence, \emph{i.e.
}the data about dynamics has to be either i.i.d. ensembles of paths or the large
collections of Markov correlated transitions in a single long path.
The former belongs to the formulation of i.i.d. ensembles and has
been reviewed before. Here, we focus on the latter generalization
where the data is Markov correlated about consequent transitions.

Let us begin by considering (a vector of) transition-based observables $\boldsymbol{g}_{ij}$
in a discrete-time Markov chain (DTMC) where $i$ and $j$ are in
the state space $\mathcal{X}$. The steady-state expected value of
$\boldsymbol{g}_{ij}$ is 
\begin{equation}
\left\langle \boldsymbol{g}\right\rangle =\sum_{i,j\in\mathcal{X}}\pi_{i}K_{j|i}\boldsymbol{g}_{ij}\label{eq: expected value of gij}
\end{equation}
where $\pi_{i}$ is the steady state distribution and $K_{j|i}$ is
the underlying transition probability matrix of the DTMC from $i$
to $j$. The ergodic theory for Markov chain tells us that the long-term
empirical average of $\boldsymbol{g}_{ij}$ converges to the steady-state
mean value $\left\langle \boldsymbol{g}\right\rangle $: 
\begin{equation}
\lim_{T\rightarrow\infty}\bar{\boldsymbol{g}}_{T}=\lim_{T\rightarrow\infty}\frac{1}{T}\sum_{t=1}^{t=T}\boldsymbol{g}_{x_{t-1},x_{t}}=\left\langle \boldsymbol{g}\right\rangle .\label{eq: transition-based data}
\end{equation}
The underlying mechanism of this convergence is the convergence of
the joint empirical frequency of a transition pair in a length-$T$
path $x_{0:T}$, 
\begin{equation}
f_{ij}(T)\coloneqq\frac{\#\text{ of }i\mapsto j\text{ in }x_{0:T}}{\text{time length }T},\label{eq: pair empirical frequency}
\end{equation}
to the steady-state pair probability $\pi_{i}K_{j|i}$ in the long-term
limit $T\rightarrow\infty$. These laws of large number for Markov
chains are the direct extension from the i.i.d. sample of distribution
to correlated-data produced by Markov processes. 

Now, similar to the i.i.d. case \citep{cheng_asymptotic_2021,lu_emergence_2022},
a key to derive MaxEnt is that the frequency $f_{ij}$ has an asymptotical
distribution with an exponential form under a prior Markov chain
model with probability $\mathbb{Q}$ \citep{barato_formal_2015}:
\begin{equation}
\mathbb{Q}\{f_{ij}\}=\exp\left[-T\sum_{i,j\in\mathcal{X}}f_{ij}\ln\frac{f_{j|i}}{R_{j|i}}+o(T)\right].\label{eq: LDT 2.5 of pair distribution}
\end{equation}
The matrix $R_{j|i}$ is our prior transition matrix that defines
our prior probability $\mathbb{Q}$, and the matrix $f_{j|i}$ is
the empirical transition matrix calculated by $f_{ij}/\sum_{k\in\mathcal{X}}f_{ik}$.
Then, we can consider three conceptually different posterior joint
stationary probability \citep{csiszar_conditional_1987}, which are
all mathematically equivalent in the long-term limit $T\rightarrow\infty:$ 

a) the asymptotic conditional probability: 
\begin{equation}
P_{ij}^{*}=\lim_{T\rightarrow\infty}\mathbb{Q}\{X_{s(T)}=i,X_{s(T)+1}=j|\bar{\boldsymbol{g}}_{T}\}\label{eq: joint pair conditioning}
\end{equation}
where the time label $s$ is a function of $T$, chosen such that
$X_{s}$ and $X_{s+1}$ are at the steady state of the process; 

b) asymptotic conditional expectation of the empirical pair frequency:
\begin{equation}
P_{ij}^{*}=\lim_{T\rightarrow\infty}\mathbb{E}[f_{ij}(T)|\bar{\boldsymbol{g}}_{T}]\label{eq: conditional expected empirical frequency}
\end{equation}
where $\mathbb{E}[\cdot]$ is taken w.r.t. the prior model $\mathbb{Q}$;

c) the most probable empirical frequency: 
\begin{align}
P_{ij}^{*} & =\arg\min_{\left\{ f\right\} }\Big\{\sum_{i,j\in\mathcal{X}}f_{ij}\ln\frac{f_{j|i}}{R_{j|i}}-\boldsymbol{\beta}\cdot(\sum_{i,j\in\mathcal{X}}f_{ij}\boldsymbol{g}_{i,j}-\bar{\boldsymbol{g}})\nonumber \\
 & -\sum_{i\in\mathcal{X}}\sigma_{i}\sum_{j\in\mathcal{X}}\left(f_{ij}-f_{ji}\right)-\nu(\sum_{i,j\in\mathcal{X}}f_{ij}-1)\Big\}.\label{eq: max ent calculation of pair}
\end{align}
The three constraints in Eq. \eqref{eq: max ent calculation of pair}
are the empirical averages of data \emph{ad infinitum}, the stationary
constraint, and the normalization. The equivalency among the three
are known as the Gibbs conditioning principle \citep{dembo_large_2009,csiszar_conditional_1987}. 

We note that the ``entropy'' to be extremized in the Markov correlated
data case here in Eq. \eqref{eq: max ent calculation of pair} is
\textit{not} the Kullback-Leibler relative entropy of pair probabilities $\sum_{i,j\in\mathcal{X}}P_{ij}\ln\frac{P_{ij}}{\pi_{i}^{R}R_{j|i}}$
where $\pi_{i}^{R}$ is the stationary distribution of prior $R_{j|i}$.
The fundamental reason of this is because MaxCal is about the whole
path $x(t),t>0$, not just one step. This can be seen from the alternative
MaxCal derivation of Eq. \eqref{eq: max ent calculation of pair}
shown in the Supplemental Material.

Since Eq. \eqref{eq: max ent calculation of pair} is a less-known
MaxEnt calculation, we briefly summarize the recipe of computing the
posterior joint probability $P_{ij}^{*}$ below. First, we construct
a tilted matrix $M_{ij}(\boldsymbol{\beta})=R_{j|i}e^{\boldsymbol{\beta}\cdot\boldsymbol{g}_{i,j}}$.
Second, we compute its largest eigenvalue $\lambda$ and the corresponding
left and right eigenvectors, $l_{i}$ and $r_{i}$ (chosen such that
$\sum_{i\in\mathcal{X}}l_{i}r_{i}=1$). The Perron-Forbenius
theorem guarantees that $\lambda$ is real and non-negative $l_{i}$
and $r_{i}$ can be found. Third, the posterior probability transition
in terms of $\boldsymbol{\beta}$ is then given by 
\begin{align}
P_{j|i}^{*}=\frac{r_{j}(\boldsymbol{\beta})}{\lambda(\boldsymbol{\beta})r_{i}(\boldsymbol{\beta})}R_{j|i}e^{\boldsymbol{\beta}\cdot\boldsymbol{g}_{i,j}}\label{eq: posterior}
\end{align}
with the stationary distribution given by $\pi_{i}^{*}=l_{i}(\boldsymbol{\beta})r_{i}(\boldsymbol{\beta})$
and $P_{ij}^{*}=\pi_{i}^{*}P_{j|i}^{*}$. Finally, we solve $\boldsymbol{\beta}$($\left\langle \boldsymbol{g}\right\rangle )$
according to $\left\langle \boldsymbol{g}\right\rangle =\nabla\ln\lambda(\boldsymbol{\beta})$,
which is a set of PDEs that can be solved systematically with optimization
procedures described later in Eq. \eqref{eq: GE theorem} and Eq.
\eqref{eq: derivative relations}.

\paragraph{Thermodynamic structures emerge from limit theorems.}

Statistical thermodynamics can be derived generally by MaxEnt \citep{presse_principles_2013,lu_emergence_2022}.
And, based on the data-driven empirical derivation of MaxEnt reviewed
above, we can consider thermodynamics as emerged from the data infinitus
limit, for both i.i.d. ensembles and Markov correlated transitions. 

The origin of the thermodynamic structure is the convex duality between
a pair of functions, known as \emph{entropy} and \emph{free energy}
in classical thermodynamics. For i.i.d. data about the distribution
of states, the ``entropy'' is the posterior relative entropy, $\varphi(\left\langle \boldsymbol{g}\right\rangle )\coloneqq\sum_{i\in\Omega}p_{i}^{*}\log\frac{p_{i}^{*}}{q_{i}}$,
and the ``free energy'' is the generating function of the observable
$\boldsymbol{g}$, $\psi(\boldsymbol{\beta})\coloneqq\log\sum_{i\in\Omega}e^{\boldsymbol{\beta}\cdot\boldsymbol{g}}$.
This is the textbook classical thermodynamics \cite{qian_counting_2020,lu_emergence_2022}. For transition-based
Markov correlated data, the ``entropy'' becomes the posterior path
relative entropy, 
\begin{equation}
\varphi(\left\langle \boldsymbol{g}\right\rangle )\coloneqq\sum_{i,j\in\mathcal{X}}P_{ij}^{*}\log\frac{P_{j|i}^{*}}{R_{j|i}},\label{eq: entropy}
\end{equation}
and the ``free energy'' is the scaled generating function for the
empirical sum $\boldsymbol{G}_{T}\coloneqq\sum_{t=1}^{T}\boldsymbol{g}_{x_{t-1},x_{t}},$
\begin{equation}
\psi\left(\boldsymbol{\beta}\right)\coloneqq\lim_{T\rightarrow\infty}\frac{1}{T}\log\mathbb{E}\left[e^{\boldsymbol{\beta}\cdot\boldsymbol{G}_{T}}\right]=\log\lambda(\boldsymbol{\beta}),\label{eq: free energy}
\end{equation}
which becomes the logarithm of the largest eigenvalue $\lambda$ computed
by the tilted matrix. In both cases, the ``free energy'' $\psi$
is a generating function, and the entropy $\varphi$ is the extremized
value of a entropy function.

Convex duality between the entropy $\varphi$ and the free energy
$\psi$ emerges from limit theorems. On the one hand, the free energy
$\psi$ is always the Legendre-Fenchel transform of the entropy $\varphi$.
\begin{equation}
\psi(\boldsymbol{\beta})=\max_{\boldsymbol{x} }\left[\boldsymbol{\beta}\cdot \boldsymbol{x} -\varphi(\boldsymbol{x})\right].\label{eq: free energy is LFT of entropy}
\end{equation}
This is a direct consequence of computing $\psi$ from its definition
with the Laplace's approximation in asymptotic analysis \citep{touchette_large_2009,qian_counting_2020}.
On the other hand, the inverse of Eq. \eqref{eq: free energy is LFT of entropy}
requires the existence and differentiability of $\psi$. This is known
as the G\"{a}rtner-Ellis theorem \citep{gartner_large_1977,ellis_large_1984,touchette_large_2009}:
\begin{align}
\varphi(\left\langle \boldsymbol{g}\right\rangle ) & =\max_{\boldsymbol{\xi}}\left[\boldsymbol{\xi}\cdot\left\langle \boldsymbol{g}\right\rangle -\psi(\left\langle \boldsymbol{\xi}\right\rangle )\right].\label{eq: GE theorem}
\end{align}
 These Legendre-Fenchel-transform expressions of $\varphi$ and $\psi$
tell us that both of them are convex functions \citep{touchette_large_2009}. With differentiable
$\psi$ and $\varphi$, the two Legendre-Fenchel transforms above
reduces to a single Legendre transform, which encodes derivative relations
between the dual coordinates $\left\langle \boldsymbol{g}\right\rangle $
and $\boldsymbol{\beta}$ of the system, 
\begin{equation}
\boldsymbol{\beta}=\nabla\varphi(\left\langle \boldsymbol{g}\right\rangle )\text{ and }\left\langle \boldsymbol{g}\right\rangle =\nabla\psi(\boldsymbol{\beta}),\label{eq: derivative relations}
\end{equation}
as well as the Maxwell's relations associated with them. Importantly,
Eq. \eqref{eq: derivative relations} shows that the parameters $\boldsymbol{\beta}$
are the \emph{entropy forces }conjugated to the observables $\left\langle \boldsymbol{g}\right\rangle $.

\paragraph{Statistical Uncertainty Principle (SUP)}

We are now ready to discuss the dynamic extension of the uncertainty
principle between the statistical variations of observables (\emph{e.g.
}energy) and of the \emph{inferred} conjugated entropic forces (\emph{e.g.
}1/temperature) in thermodynamics \citep{landau_statistical_1980,mandelbrot_temperature_1989,uffink_thermodynamic_1999,schlogl_thermodynamic_1988}.
We shall call it the statistical uncertainty principle (SUP) since
it is an leading-order statistical result for large but finite data.
Our contributions here are twofold: a) We extends SUP
from state observables \citep{landau_statistical_1980,mandelbrot_temperature_1989,uffink_thermodynamic_1999}
to transition observables, from distribution to dynamics; b) SUP shows
the physical meaning of a well-known mathematical relation in large
deviation theory.

To illustrate, let's consider the following scenario. Suppose Bob
has transition-based data in the form of the empirical mean $\bar{\boldsymbol{g}}_{T}=\sum_{i,j\in\mathcal{X}}f_{T}(i,j)\boldsymbol{g}(i,j)$
for a large but finite data of length $T$. Note that $\bar{\boldsymbol{g}}_{T}$
is itself random: if Bob repeats the experiment, he can get different
values of $\bar{\boldsymbol{g}}_{T}$. Now, suppose Alice knows the
true expected values $\left\langle \boldsymbol{g}\right\rangle $
of the transition observable $\boldsymbol{g}(i,j)$ either because
she had data \emph{ad infinitum} or due to other sources, she can
then predict the asymptotic variation of Bob's $\bar{\boldsymbol{g}}_{T}$
quantified by the covariance matrix of Bob's $\bar{\boldsymbol{g}}_{T}$
to the leading order, 
\begin{equation}
{\rm Co}\mathbb{V}[\bar{\boldsymbol{g}}_{T}]\sim\frac{1}{T}\nabla\nabla\psi\left(\boldsymbol{\beta}\right).\label{eq: covariance of bar g}
\end{equation}
Bob can verify this by repeatedly measuring $\bar{\boldsymbol{g}}_{T}$
from i.i.d. copies of the length-$T$ process.

Each time Bob gets a $\bar{\boldsymbol{g}}_{T}$, he can use it to
\emph{infer} the level of entropic forces $\boldsymbol{\beta}$. He
computes $\bar{\boldsymbol{\beta}}_{T}=\nabla\varphi(\bar{\boldsymbol{g}}_{T})$
by using the entropy function $\varphi$ given by Eq. \eqref{eq: entropy}
and plug in the $\bar{\boldsymbol{g}}_{T}$ he measured. This inferred
force $\bar{\boldsymbol{\beta}}_{T}$ fluctuates due to the stochasticity
of $\bar{\boldsymbol{g}}_{T}$. Alice can also derive the leading-order
fluctuation of Bob's $\bar{\boldsymbol{\beta}}_{T}$, which becomes
\begin{equation}
{\rm Co}\mathbb{V}[\bar{\boldsymbol{\beta}}_{T}]\sim\frac{1}{T}\nabla\nabla\varphi\left(\left\langle \boldsymbol{g}\right\rangle \right).\label{eq: covariance of the inferred forces}
\end{equation}
See Supplemental Material for a derivation. Since $\varphi$ and $\psi$
have a reciprocal curvature due to the Legendre transform \citep{touchette_large_2009},
we then have the SUP for the fluctuations of $\bar{\boldsymbol{g}}_{T}$
and of the $\bar{\boldsymbol{\beta}}_{T}$ they infer: 
\begin{equation}
\underset{\sim T{\rm Co}\mathbb{V}[\bar{\boldsymbol{g}}_{T}]}{\underbrace{\nabla\nabla\psi\left(\boldsymbol{\beta}\right)}}\underset{\sim T{\rm Co}\mathbb{V}[\bar{\boldsymbol{\beta}}_{T}]}{\underbrace{\nabla\nabla\varphi\left(\left\langle \boldsymbol{g}\right\rangle \right)}}=\mathbf{I}.\label{eq: ATUP}
\end{equation}
While this mathematical relation is well-known to the large deviation
theory community \citep{touchette_large_2009}, to our knowledge this
is the first time its statistical meaning is pointed out. It is worth
noticing that Schl\"{o}gl has derived an inequality version of SUP without
taking data infinitus limit, which can be regarded as the mesoscopic
origin of SUP \citep{schlogl_thermodynamic_1988}.

\paragraph{A simple toy model of molecular motor as an example}

\begin{figure}
\begin{centering}
\includegraphics[width=0.65\columnwidth]{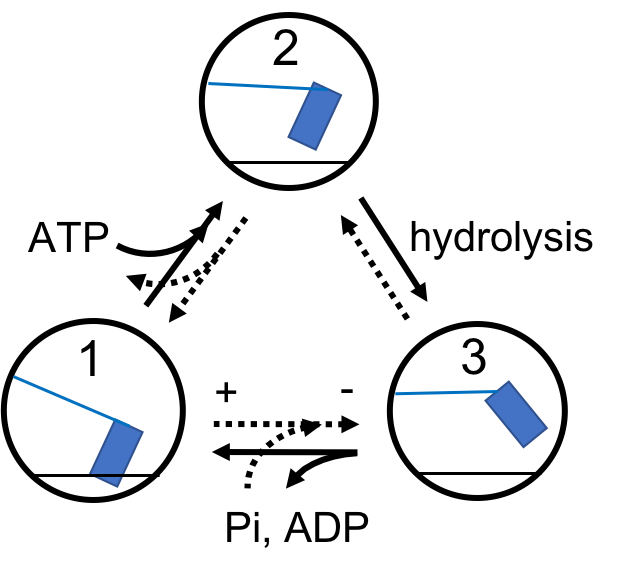}
\par\end{centering}
\caption{The state space of a three-state toy model of molecular motor (myosin).
Solid (dashed) arrows indicate transitions in a forward (backward)
cycle.\label{fig: state-space}}
\end{figure}

To illustrate SUP, let us consider a simple three-state Markov chain
as a toy model for a molecular motor monomer like myosin \citep{qian_simple_1997}.
Our toy molecular motor is assumed to have three states with state
space illustrated in Fig. \ref{fig: state-space}. 
 The motor at state 1 is bounded to the actin. Through coupling with
an ATP, it detaches and becomes state 2. Then, hydrolysis of ATP leads
to the mechanically deformed state 3. Through releasing ADP and Pi,
the motor generates a power stroke (from - to +) and re-attach to
the actin, back to state 1. The dynamics of this motor (in discrete
time) is described by the transition probabilities $P_{j|i}$ from
$i$ to $j$, where $i,j\in\left\{ 1,2,3\right\} $.

With big data about a long trajectory of the motor's state, we can
use counting statistics to infer $P_{j|i}$. Let us consider the following
six linearly independent counting frequencies: the occurance frequency
of two of the three states, say frequencies $f_{2}$ and $f_{3}$,
the symmetric flux (what Maes called \emph{traffic} in \citep{maes_frenesy_2020})
over all edges measured by 
\begin{equation}
f_{ij}^{\text{sym}}=\frac{\left(\text{\# of }i\mapsto j\right)+\left(\text{\# of }j\mapsto i\right)}{\text{length of trajectory}}\label{eq: traffic}
\end{equation}
where $ij=\{12,23,13\}$, and the net (antisymmetric) flux of the
power stroke step, from state $3$ to $1$, denoted by 
\begin{equation}
f_{31}^{\text{anti}}=\frac{\left(\text{\# of }3\mapsto1\right)-\left(\text{\# of }1\mapsto3\right)}{\text{length of trajectory}}.\label{eq: net flux}
\end{equation}
Note that this $f_{31}^{{\rm anti}}$ is the net empirical velocity
of the motor over a length $T$ trajectory.

Following the MaxEnt recipe mentioned above, we assume a uniform prior
$R_{j|i}=1/3$ and construct the tilted matrix 
\begin{equation}
M_{ij}=\frac{1}{3}\left(\begin{array}{ccc}
1 & e^{\beta_{12}} & e^{\beta_{13}-\gamma}\\
e^{\alpha_{2}+\beta_{12}} & e^{\alpha_{2}} & e^{\alpha_{2}+\beta_{23}}\\
e^{\alpha_{3}+\beta_{13}+\gamma} & e^{\alpha_{3}+\beta_{23}} & e^{\alpha_{3}}
\end{array}\right)\label{eq:tilted matrix}
\end{equation}
with six parameters $(\alpha_{2},\alpha_{3},\beta_{12},\beta_{23},\beta_{13},\gamma)$
corresponding to the six observables $\boldsymbol{f}=\left(f_{2},f_{3},f_{12}^{{\rm sym}},f_{23}^{{\rm sym}},f_{13}^{{\rm sym}},f_{31}^{{\rm anti}}\right)$.
Then by Eq. \eqref{eq: posterior}, the posterior transition probabilities
take the form of 
\begin{equation}
P_{j|i}^{*}=\frac{1}{3\lambda}\left(\begin{array}{ccc}
1 & e^{\beta_{12}}\frac{r_{2}}{r_{1}} & e^{\beta_{13}-\gamma}\frac{r_{3}}{r_{1}}\\
e^{\alpha_{2}+\beta_{12}}\frac{r_{1}}{r_{2}} & e^{\alpha_{2}} & e^{\alpha_{2}+\beta_{23}}\frac{r_{3}}{r_{2}}\\
e^{\alpha_{3}+\beta_{13}+\gamma}\frac{r_{1}}{r_{3}} & e^{\alpha_{3}+\beta_{23}}\frac{r_{2}}{r_{3}} & e^{\alpha_{3}}
\end{array}\right)\label{eq: posterior transition probability}
\end{equation}
where $\lambda$ is the largest eigenvalue of $M$ and $r$ is the
corresponding right eigenvector.

The set of observables $\boldsymbol{f}$ we chose is \emph{holographic}, \emph{i.e.
}it captures all degrees of freedom of the dynamics. When the trajectory
 becomes very long, the ergodic theorem of Markov chain guarantees
that $\left(f_{2},f_{3}\right)\rightarrow\left(\pi_{2},\pi_{3}\right)$,
\begin{equation}
f_{ij}^{{\rm sym}}\rightarrow\tau_{ij}=P_{ij}+P_{ji},\label{eq: traffic convergence}
\end{equation}
and 
\begin{equation}
f_{31}^{{\rm anti}}\rightarrow J=P_{31}-P_{13}.\label{eq: net flux convergence}
\end{equation}
With these six averages $\left\langle \boldsymbol{f}\right\rangle =(\pi_{2},\pi_{3},\tau_{12},\tau_{23},\tau_{13},J)$,
one can uniquely compute the true underlying $P_{j|i}$ as a function
of these six averages. Furthermore, since our observables are \emph{non-degenerate},
simple relations between the six parameters and the transition probabilities
can be derived \footnote{One of us is writing a paper regarding the general version of this
fact.}:\begin{subequations}\label{eqs: entropic force}
\begin{align}
\alpha_{n} & =\ln P_{n|n}-\ln P_{1|1}\label{eq:}\\
\beta_{ij} & =\frac{1}{2}\ln\frac{P_{j|i}P_{i|j}}{P_{i|i}P_{j|j}}\label{eq:-6}\\
\beta_{12} & =\frac{1}{2}\ln\frac{P_{2|1}P_{1|2}}{P_{1|1}P_{2|2}}\label{eq:-7}\\
\gamma & =\frac{1}{2}\log\frac{P_{2|1}P_{3|2}P_{1|3}}{P_{3|1}P_{2|3}P_{1|2}}\label{eq:-9}
\end{align}
\end{subequations}where $n=\{2,3\}$ and $ij=\{12,23,13\}$. Notice
that $\text{\ensuremath{\gamma}}$ is (half of) the cycle affinity,
an important term in stochastic thermodynamics \citep{yang_bivectorial_2021}.

Recall from Eq. \eqref{eq: derivative relations} that the six parameters
in Eqs. \eqref{eqs: entropic force} are actually the entropic forces.
In our example here, we plug in $R_{j|i}=1/3$ and 
\begin{equation}
P_{j|i}^{*}=\frac{\tau_{ij}+J_{ij}}{2\pi_{i}}\label{eq: posterior P_j|i}
\end{equation}
into the entropy form $\varphi(\left\langle \boldsymbol{f}\right\rangle )$
in Eq. \eqref{eq: entropy} by using normalization $\pi_{1}=1-\pi_{2}-\pi_{3}$
and stationarity $J_{12}=J_{23}=J_{31}=J$. One can easily check that
\begin{equation}
\alpha_{n}=\frac{\partial\varphi}{\partial\pi_{n}},\beta_{ij}=\frac{\partial\varphi}{\partial\tau_{ij}},\gamma=\frac{\partial\varphi}{\partial J}\label{eq: entropic forces}
\end{equation}
for $n=2,3$ and $ij\in\{12,23,31\}$.

\begin{figure}
\begin{centering}
\includegraphics[width=1\columnwidth]{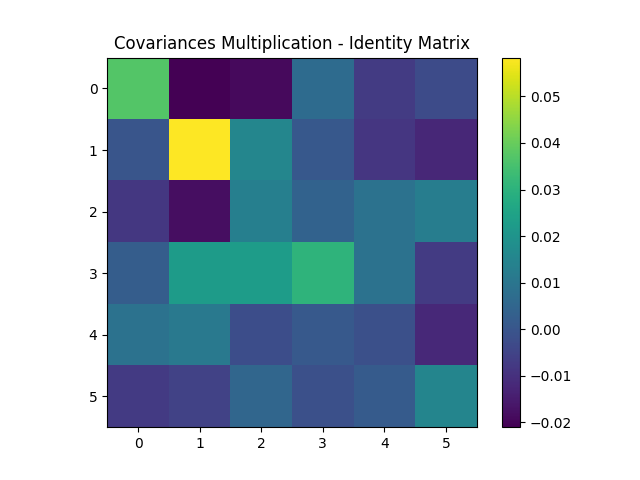}
\par\end{centering}

\caption{The image shows the difference between the product of statistic variations
of frequencies and forces, $T^{2}{\rm Co}\mathbb{V}[\boldsymbol{f}]{\rm Co}\mathbb{V}[\boldsymbol{F}]$,
in a simulated data and the identify matrix. The underlying transition
probabilities are $P_{:|1}=\left(\frac{13}{40},\frac{9}{20},\frac{9}{20}\right),P_{:|2}=\left(\frac{3}{20},\frac{1}{4},\frac{3}{5}\right),$
$P_{:|3}=\left(\frac{21}{40},\frac{3}{10},\frac{7}{40}\right)$. This
is chosen arbitrarily so that $\left(\pi_{2},\pi_{3},\tau_{12},\tau_{23},\tau_{13},J\right)=\left(\frac{1}{3},\frac{1}{3},\frac{1}{5},\frac{1}{4},\frac{3}{10},\frac{1}{10}\right)$.
We simulate $M=10^{5}$ i.i.d. copies of length $T=1000$ trajectories and
compute the six empirical frequencies and the inferred forces from
each trajectories. The differences would shrink to zero when $M\rightarrow\infty$
and $T\rightarrow\infty$.\label{fig: cov cov -I}}

\end{figure}

The SUP in Eq. \eqref{eq: ATUP} is an asymptotic relation between
the covariance of the six frequency observables $\boldsymbol{f}$
collected from a long but finite trajectory and the forces they inferred
by computing $\boldsymbol{F}=\nabla\varphi(\boldsymbol{f}).$ By numerically
produce a big ensemble of very long trajectories with length $T$,
we can play Bob's role and check the SUP for the empirical $6\times6$ scaled covariance of
frequencies $T{\rm Co\mathbb{V}[\text{\ensuremath{\boldsymbol{f}}}]}$
and that of the inferred forces $T{\rm Co}\mathbb{V}[\boldsymbol{F}]$.
Their product is indeed very close to the identify matrix with vanishing
difference to the identity matrix shown in Fig. \eqref{fig: cov cov -I}.

\paragraph{Summary \label{sec:Conclusion-and-Discussions}}

In this paper, we revisit and advocate the empirical derivation of
the Maximum Entropy principle of Markov correlated data \citep{csiszar_conditional_1987,chetrite_nonequilibrium_2013,chetrite_nonequilibrium_2015},
also known as the Maximum Caliber principle. We review how the principle
can identify entropic forces, and lead to statistical thermodynamic
conjugacy. From the empirical understanding and data-driven derivation that we revisited and advocated, we derived an uncertainty
principle between the statistical variations of observables and the
forces they infer from finite data. This theory is purely empirical
and can thus be applied to trajectory data from small, nonequilibrium, dynamical biological systems that are far from mechanics. 

In short, more is indeed
different \citep{anderson_more_1972}. Maximum entropy principle and
statistical thermodynamics emerge empirically and \emph{de-mechanically}
from the limit theorems of data \emph{ad infinitum}, i.i.d. or Markov correlated.
Akin to the uncertainty principle in quantum mechanics, there is an uncertainty principle about the statistical variations of dynamical observables and forces for Big Data.

\begin{acknowledgments}
The authors thank Erin Angelini, Ken Dill, Charles Kocher, Zhiyue Lu, Jonny Patcher, Dalton Sakthivadivel,
Dominic Skinner, David Sivak, Lowell Thompson, and Jin Wang for helpful
feedback on our manuscript and for the stimulating discussions they
had with the authors. H. Q. thanks the support from the Olga Jung
Wan Endowed Professorship.
\end{acknowledgments}

\bibliography{ST-Data-Infinitum}

\end{document}